\documentclass[prx,twocolumn,preprintnumbers,amsmath,amssymb]{revtex4}

\usepackage{graphicx}
\usepackage{dcolumn}
\usepackage{bm}
\usepackage{mathrsfs}
\usepackage{subfigure}
\usepackage{natbib}
\usepackage{amsmath}
\usepackage{amssymb}
\usepackage{Jonasmacros}
\usepackage[usenames,dvipsnames,svgnames]{xcolor}

\begin{document}

	\title{Dynamical exchange and phase induced switching of a localized molecular spin}

	\author{H. Hammar}
	\affiliation{Department of Physics and Astronomy, Uppsala University, Box 530, SE-751 21 Uppsala}

	\author{J. Fransson}
	\affiliation{Department of Physics and Astronomy, Uppsala University, Box 530, SE-751 21 Uppsala}

	\date{\today}

\begin{abstract}
We address the dynamics of a localized molecular spin under the influence of external voltage pulses using a generalized spin equation of motion which incorporates anisotropic fields, nonequilibrium conditions, and non-adiabatic dynamics. We predict a recurring $4\pi$-periodic switching of the localized spin by application of a voltage pulse of temporal length $\tau$. The switching phenomena can be explained by dynamical exchange interactions, internal transient fields, and self-interactions acting on the localized spin moment. 
\end{abstract}

\maketitle

\section{Introduction}
Dynamics of open systems is an active area of research \cite{DeVega2017,Breuer2015a}. Recent theoretical predictions have suggested that periodical out-of-equilibrium driving can induce temporal phases of matter \cite{Wilczek2012}, which subsequently have been experimentally corroborated \cite{Yao2017,Choi2017}.
Light induced ultra-fast demagnetization has shown that fast responses to external forces can change the long-term magnetic properties, approaching stationary regimes not accessible through adiabatic processes \cite{Walowski2016}. These examples vividly illustrate that the equilibrium paradigm is insufficient when attempting to treat rapid dynamics and nonequilibrium systems.
Thus, when approaching the quantum limit in both spatial and temporal dimensions, models based on instantaneous or local interactions with no record of the past evolution or spatial surrounding can always be questioned. Non-linearities and feedback between internal components require a higher level of sophistication in the theoretical modeling, allowing to go beyond the equilibrium narrative, especially when confinement plays an important role as in single molecules.

Nonequilibrium open systems such as nanojunctions, quantum dots, and single molecules have been studied extensively, both experimentally and theoretically. Studies include electron dynamics \cite{Kurzmann2017,Roche2013}, vibrating quantum dots \cite{Fransson2010c}, pulse-enhanced thermoelectric efficiency \cite{Crepieux2011, Zhou2015}, nonequilibrium thermodynamics \cite{Esposito2015,Esposito2015b}, and optoelectronics and spectroscopy \cite{Galperin2017,Grosse2013}. Due to size confinement, the systems exhibit intrinsic out-of-equilibrium nature and can be controlled by pulses and external forces, thus well suited for studying non-adiabatic quantum dynamics.

\begin{figure}[b]
	\centering
	\includegraphics[width=\columnwidth]{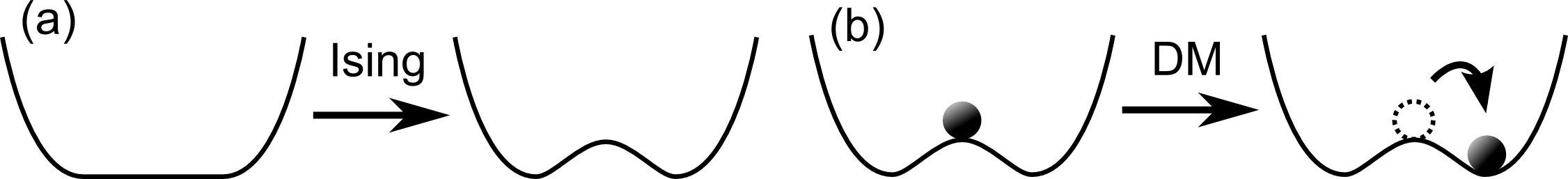}
	\caption{Illustration of the contribution of (a) the Ising interaction/uniaxial anisotropy and (b) the DM interaction. (a) The Heisenberg interaction supports two degenerate solutions for the local spin for which there is no energy barrier in between. The Ising interaction and intrinsic uniaxial anisotropy introduces a potential barrier, and by that creating two separate minima. (b) The DM interaction provides the mechanism of the spin to switch and fall into the potential wells.}
	\label{potentials}
\end{figure}

In this article we predict a novel type of phase induced switching phenomenon of localized spin embedded in a tunnel junction between metallic leads, across which a time-dependent voltage, $V(t)$, is applied. By application of a voltage pulse of temporal length $\tau$, we observe a recurring switching property of the localized spin, essentially whenever the total accumulated phase $\varphi(V,\tau)\equiv eV\tau/\hbar\in(2\pi,4\pi) \mod{4\pi}$.
The build up of the accumulated phase generates highly anisotropic internal transient fields which act on the local spin, exerting a torque which counteracts the externally applied magnetic field. The altered spin configuration is stabilized by an intrinsic uniaxial anisotropy field of the localized spin and the internal fields crucially governs the dynamics long after the voltage pulse is turned off.
This novel switching phenomenon can be explained in terms of induced internal transient fields emerging during the voltage pulse. These can be partitioned into four components: i) internal magnetic field, ii) Heisenberg, iii) Dzyaloshinskii-Moriya (DM), and iv) Ising type of self-interactions between the spin at different times. 
While all four components are essential for the switching, we notice in particular that the intrinsic uniaxial anisotropy and the dynamic Ising interaction creates an energy barrier between degenerate solutions for the spin, see Fig. \ref{potentials}(a), which is crucial to stabilize the steady state after switching, whereas the DM interaction provides a torque that is required to drive the spin out of its initial state into the a new final state, see Fig. \ref{potentials}(b). The switching depends heavily on the sign of the DM interaction which can be controlled by tuning the intrinsic uniaxial anisotropy, exchange coupling, temperature and external magnetic field.

Our results are obtained from a generalized spin equation of motion (SEOM) developed for nonequilibrium conditions \cite{Fransson2008a,Fransson2008b,Diaz2012,Nunez2008,Katsura2006} and which allows for calculations of dynamic exchange interactions \cite{Zhu2004,Fransson2014a,Bhattacharjee2012,Fransson2010,Fransson2008a}. Similar approaches have previously been used in the stationary limit \cite{Duine2007,Shnirman2015,Onoda2006,Katsura2006,Saygun2016a}. In comparison to previous studies using, e.g., quantum master equation \cite{Misiorny2007,Metelmann2012,Mosshammer2014} and stochastic Landau-Lifshitz-Gilbert equation \cite{Filipovic2013,Bode2012}, our approach makes a full account of the non-adiabatic dynamics, including temporal non-local properties of the internal fields. This has shown to be of great importance in studies of, e.g., ultra-fast spin dynamics \cite{Secchi2013,Secchi2016,Mikhaylovskiy:2015aa,Mueller2013,Mentink2014}.

Our test bench model represents a single-molecule magnet, for instance $M$-porphyrins and $M$-phthalocyanines where $M$ denotes, e.g., a transition metal element, which serve as good models for fundamental studies \cite{Bogani:2008aa,Locatelli:2014aa,Chappert:2007aa} comprising an inherent nonequilibrium nature.
Experiments have revealed distance dependent effects in the exchange interactions \cite{Hirjibehedin2006,Wahl2007,Zhou:2010aa,Meier2008},
large anisotropy of individual molecules  \cite{Rau2014,Bairagi2015,Voss2008,Balashov2009}, as well as collective spin excitations and Kondo effect \cite{Chen2008,Pruser:2011aa,Khajetoorians2013}.
Experiments have also shown the control and read-out of spin states of individual single-molecule magnets \cite{Hauptmann:2008aa,Loth:2010aa,Wagner:2013aa, Vincent2012,Loth2010b,Ternes2015,Loth2010,Karan2016,Lin2015}.

\section{Model}
We consider a magnetic molecule, embedded in a tunnel junction between metallic leads, comprising a localized magnetic moment ${\bf S}$ coupled via exchange to the highest occupied or lowest unoccupied molecular orbital henceforth referred to as the QD level. We define our system Hamiltonian as
\begin{equation}
{\cal H}
=
{\cal H_{{\rm \chi}}}+{\cal H_{{\rm T}}}+{\cal H_{{\rm QD}}}+{\cal H_{{\rm S}}}
.
\end{equation}
Here, ${\cal H}_{\chi}=\sum_{\bfk\sigma\in\chi}(\varepsilon_{\bfk\chi}-\mu_{\chi})c_{\bfk\chi\sigma}^{\text{\ensuremath{\dagger}}}c_{\bfk\chi\sigma},$
is the Hamiltonian for the left ($\chi=L$) or right ($\chi=R$) lead, where $c_{\bfk\chi\sigma}^{\dagger}$ ($c_{\bfk\chi\sigma}$) creates (annihilates) an electron in the lead $\chi$ with energy $\varepsilon_{\bfk\chi}$, momentum \textbf{k}, and spin $\sigma=\up,\down$, while $\mu_{\chi}$ denote the chemical potential such that the voltage $V$ across the junction is defined by $eV = \mu_{L}-\mu_{R}$.
Tunneling between the leads and the QD level is described by ${\cal H}_{T}={\cal H}_{TL}+{\cal H}_{TR}$, where $
{\cal H}_{T\chi}=T_{\chi}\sum_{\bfk\sigma\in\chi}c_{\mathbf{k\chi\sigma}}^{\dagger}d_{\sigma}+H.c.$.
The single-level QD is represented by ${\cal H}_{QD}=\sum_{\sigma}\varepsilon_{\sigma}d_{\sigma}^{\text{\ensuremath{\dagger}}}d_{\sigma}$, where $d_{\sigma}^{\dagger}$ ($d_{\sigma}$) creates (annihilates) an electron in the QD with energy $\varepsilon_{\sigma} = \varepsilon_{0} + g\mu_{B} B^\text{ext} \sigma^{z}_{\sigma\sigma}/2$ and spin $\sigma$, depending on the external magnetic field $\bfB^\text{ext}=B^\text{ext}\hat{\bf z}$, where g is the gyromagnetic ratio and $\mu_{B}$ the Bohr magneton. 
The energy of the local spin is described by ${\cal H}_{\rm S}=-g\mu_B\bfS\cdot\bfB^\text{ext}-v\mathbf{s}\cdot\mathbf{S}+DS_{z}^2$ where $v$ is the exchange integral between the localized and delocalized electrons, the electron spin is denoted \textbf{$\mathbf{s}=\psi^{\dagger}\boldsymbol{\sigma}\psi/2$} in terms of the spinor $\psi=(d_\up\ d_\down)$, $\boldsymbol{\sigma}$ is the vector of Pauli matrices and D is an intrinsic uniaxial anisotropy field in the magnetic molecule.

The local spin dynamics is calculated using our previously developed generalized SEOM \cite{Hammar2016}, that is,
\begin{align}
	\dot{\mathbf{S}}(t) = &\mathbf{S}(t)\times\left( -g\mu_{B}\mathbf{B}^{\mathrm{eff}}_{0}(t) \right.
	\nonumber
	\\
	&\left.+\frac{1}{e}\int (\mathbb{J}(t,t')+\mathbb{D})\cdot\mathbf{S}(t')dt'\right).
	\label{spinequationofmotion}
\end{align}
Here, $\mathbf{B}^\text{eff}_{0}(t)$ is the effective magnetic field acting on the spin, defined by $\mathbf{B}^{\mathrm{eff}}_{0}(t)=\textbf{B}^{\mathrm{ext}}+\frac{v}{g\mu_{B}}\bfm(t)-\int\textbf{j}(t,t')dt'/eg\mu_{B}$,
where the second contribution is the local electronic magnetic moment, defined as $\bfm (t) = \left\langle \mathbf{s}(t) \right\rangle = \frac{1}{2}\left\langle \psi(t)^\dagger \bfsigma \psi(t)\right\rangle = \frac{1}{2}\text{Imsp}{\bfsigma}\textbf{G}^<(t,t)$, where ${\rm sp}$ denotes the trace over spin-1/2 space. The third term is the internal magnetic field due to the electron flow.  The field $\mathbb{J}(t,t')$ is the dynamical exchange coupling between spins at different times and $\mathbb{D} = D \mathbf{\hat{z}\hat{z}}$ is due to the intrinsic uniaxial anisotropy.

The generalized SEOM makes use of the Born-Oppenheimer approximation which is motivated as the energy scales of single molecule magnets are in meV which results in spin dynamics of picoseconds. This is orders of magnitudes smaller than the recombination time-scales of the electrons in the junction in the orders of femtoseconds. We also remark that despite the semi-classical nature of the generalized SEOM, it incorporates the underlying quantum nature of the junction through the dynamical fields ${\bf j}$ and $\mathbb{J}$. This is especially important in the transient regime, where the classical Landau-Lifshitz-Gilbert equation is incapable to provide an adequate description of the dynamics \cite{Hammar2017}. The treatment goes beyond the adiabatic limit considered in previous works, e.g., Ref. \cite{Bode2012}, while still containing important attributes as dissipative fields and spin-transfer torques.

The internal magnetic field due to the electron flow is defined as
$\textbf{j}(t,t')=iev\theta(t-t')\av{\com{s^{(0)}(t)}{\mathbf{s}(t')}}$, where the on-site energy distribution is represented by $\text{s}^{(0)}=\sum_{\sigma}\varepsilon_{\sigma}d_{\sigma}^{\dagger}d_{\sigma}$. This two-electron propagator $\textbf{j}(t,t')$ is approximated by decoupling into single electron nonequilibrium Green functions (GFs), ${\bf G}^{</>}$, according to
\begin{align}
\mathbf{j}(t,t')\approx&
iev\theta(t-t'){\rm sp}\boldsymbol{\epsilon}
\Bigl(
\mathbf{G}^{<}(t',t)\mathbf{\boldsymbol{\sigma}G}^{>}(t,t')
\nonumber
\\
&
-\mathbf{G}^{>}(t',t)\mathbf{\boldsymbol{\sigma}G}^{<}(t,t')
\Bigr),
\label{currentMF}
\end{align}
where $\boldsymbol{\epsilon}={\rm diag}\lbrace \varepsilon_\uparrow\ \varepsilon_\downarrow\rbrace$. This internal field mediates both the magnetic field generated by the charge flow as well as the effect of the external magnetic field causing the Zeeman split in the QD.

The spin susceptibility tensor $\mathbb{J}(t,t')=i2ev^2\theta(t-t')\av{\com{\bfs(t)}{\bfs(t')}}$ mediates the interactions between the localized magnetic moment at the times $t$ and $t'$. Decoupling into single electron GFs, yields
\begin{align}
\mathbb{J}(t,t')\approx&
\frac{ie}{2}v^{2}\theta(t-t'){\rm sp}\boldsymbol{\sigma}
\Bigl(
\mathbf{G}^{<}(t',t)\mathbf{\boldsymbol{\sigma}G}^{>}(t,t')
\nonumber
\\
&
-\mathbf{G}^{>}(t',t)\mathbf{\boldsymbol{\sigma}G}^{<}(t,t')
\Bigr).
\label{spinsusceptibilty}
\end{align}
This current mediated interaction can be decomposed into an isotropic Heisenberg interaction $J_H$, and the anisotropic Dzyaloshinski-Moriya (DM) ${\bf D}$ and Ising $\mathbb{I}$ interactions \cite{Fransson2014a,Hammar2016}.

The dynamical QD electronic structure is calculated by using nonequilibrium GFs taking into account the back action from the local spin dynamics by perturbation theory. Expanding the contour ordered single electron GF $G(t,t')$ to first order in the time-dependent expectation value of the spin, we obtain
\begin{align}
\mathbf{G}(t,t')=&
{\bf g}(t,t')-v\oint_C {\bf g}(t,\tau)\left\langle \mathbf{S}(\tau)\right\rangle \mathbf{\cdot\boldsymbol{\sigma}}{\bf g}(\tau,t')d\tau.
\label{GF}
\end{align}
Here, ${\bf g}(t,t')$ is the bare (spin-dependent) QD GF given by the equation of motion
\begin{align}
(i\partial_{t}-\boldsymbol{\epsilon}){\bf g}(t,t')=&\delta(t-t')\sigma^0+\int\boldsymbol{\Sigma}(t,\tau){\bf g}(\tau,t')d\tau
,
\end{align}
where the self-energy is $\boldsymbol{\Sigma}(t,t')=\Sigma(t,t')\sigma^0$ with $\Sigma(t,t')=\sum_\chi\sum_{\bfk\in\chi}|T_\chi|^2g_{\bfk}(t,t')$, $\sigma^0$ is the $2\times2$ identity matrix and $g_{\bfk}(t,t')$ is the lead GF.
Using the wide-band limit we can define the tunneling coupling $\Gamma^\chi=2\pi|T_\chi|^2\sum_{\bfk\in\chi}\delta(\omega-\varepsilon_{\bfk})$ between the lead and the QD and the
lesser self-energy becomes
\begin{align}
\Sigma^{<}(t,t')=&
i\sum_\chi\Gamma^\chi \int f_{\chi}(\omega)e^{-i\omega(t-t')+i\int_{t'}^{t}\mu_{\chi}(\tau)d\tau}\frac{d\omega}{2\pi}.
\label{selfenergy}
\end{align}
The self-energy carries the information of the pulse due to the time integration of the chemical potential for each lead, i.e., $i\int_{t'}^{t}\mu_{\chi}(\tau)d\tau$. We refer to Ref. \citenum{Hammar2016} for more details.

\begin{figure*}
	\centering
	\includegraphics[width=0.8\textwidth]{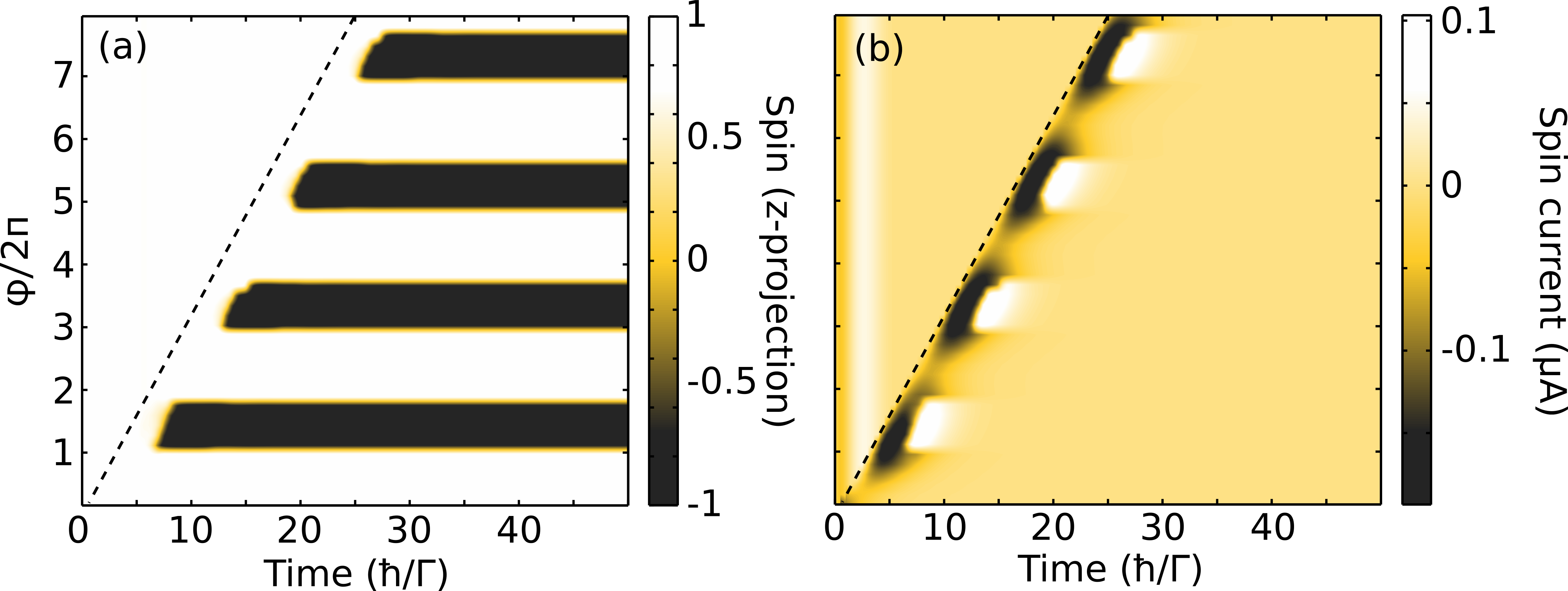}
	\caption{Resulting evolution of (a) $S_{z}$, showing the spin flip, and (b) the spin current, for different pulse lengths, plot against $\varphi/2\pi$. Here, $eV = 2\Gamma$, $v = \Gamma/2$, $D = 0.3\Gamma$, T = 0.0862 $\Gamma/\text{k}_{B}$ and B = 0.1158 $\Gamma/\text{g}\mu_{B}$. The  dotted line indicates when the pulse ends.}
	\label{ripples}
\end{figure*}

\section{Results}
In absence of a voltage across the junction, there is no current and the local spin remains in its initial state. Taking this as the initial condition for our simulations, at time $t_0$ we apply a constant voltage of amplitude $V$, which is subsequently terminated at $t_1$, and let the system evolve towards its stationary state. The plot in Fig. \ref{ripples}(a) shows the time-evolution of the local spin orientation for increasing phase $\varphi\equiv eV(t_1-t_0)/\hbar$, where bright (dark) corresponds to a spin orientation parallel (anti-parallel) to the external field. The plot clearly illustrates that the spin either remains in its initial state or is switched to the parallel state, depending on the phase. In particular for phases when $\varphi\in(0,2\pi)\mod{4\pi}$, the general orientation of the spin remains unchanged by the temporary nonequilibrium conditions while the spin aligns anti-parallel to the external field whenever $\varphi\in(2\pi, 4\pi)\mod{4\pi}$. However, due to non-linearities in Eq. (\ref{spinequationofmotion}), the two solutions are not perfectly confined to phases in the intervals $\varphi\in(0,2\pi)\mod{4\pi}$ and $\varphi\in(2\pi,4\pi)\mod{4\pi}$. We shall, nonetheless, henceforth refer to the former regime as \emph{spin-conserving} and the latter as \emph{spin-flipping}.

The spin current $I_S=\sum_\sigma\sigma_{\sigma\sigma}^zI_\sigma$, where $I_\sigma$ is the spin resolved electron current through the junction, is plotted in Fig. \ref{ripples}(b). The signatures in the spin current originates from the variations in the local spin orientation as function of the phase $\varphi$. This is expected since the spin-dependent current is sensitive to the local magnetic environment which strongly depends on whether the local spin is parallel or anti-parallel to the external magnetic field.

\begin{figure}
	\centering
	\includegraphics[width=\columnwidth]{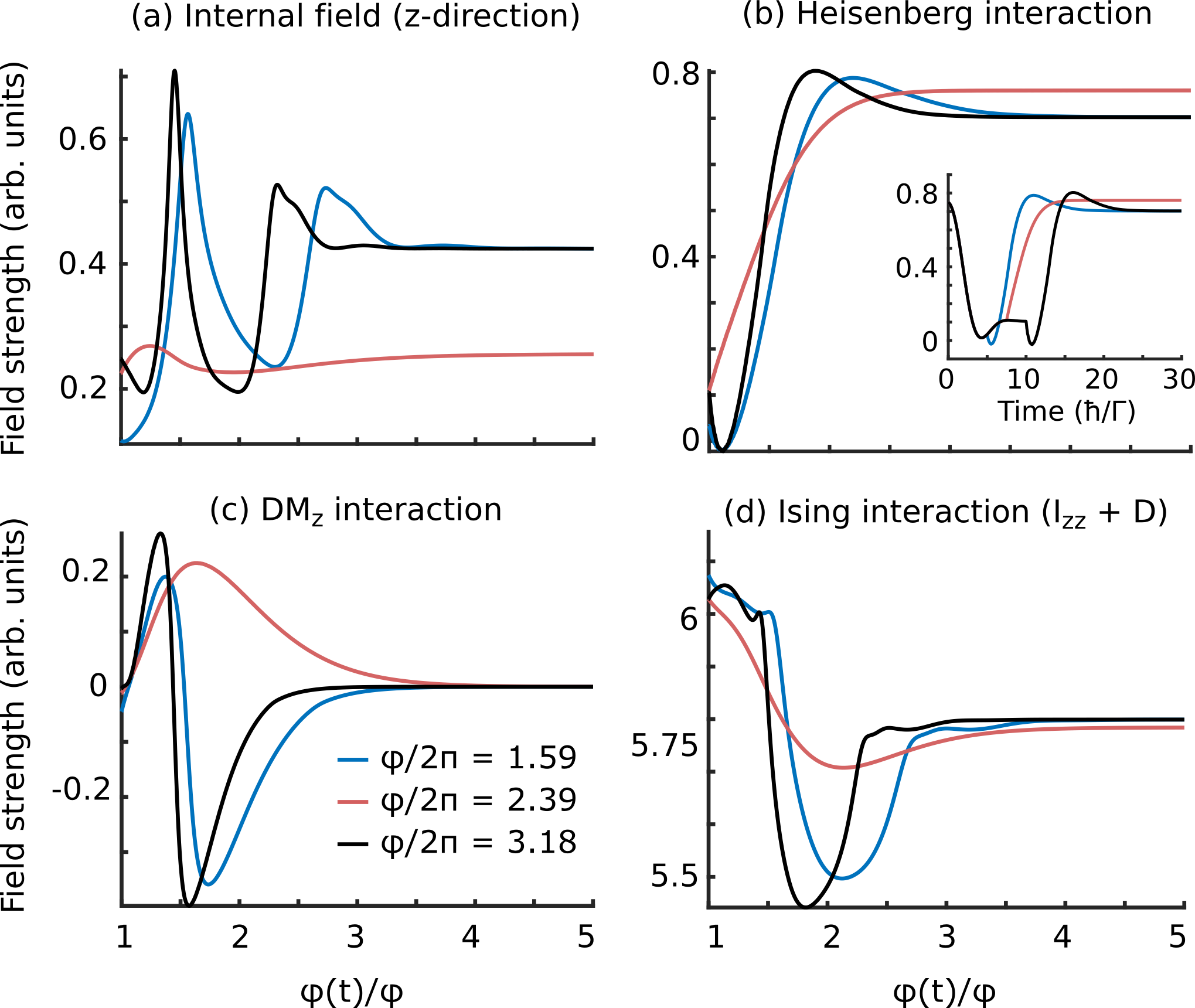}
	\caption{(a) The internal magnetic field, (b) the Heisenberg interaction, (c) the DM interaction and (d) the Ising interaction for different values of $\varphi/2\pi$. The figures plot against $\varphi(t)/\varphi$ after the pulse is turned off where $\varphi(t)=eV(t-t_0)\hbar$ and the inset in (b) show the same field against time in $\hbar/\Gamma$ for the full process. The pulse length is 5 (blue), 7.5 (red) and 10 (black) in units of $\hbar/\Gamma$. $\varphi(t)$ and $\varphi$ is in terms of mod $2\pi$. Other parameters as in Fig. \ref{ripples}.}
	\label{interactions}
\end{figure}

The origin of the phase induced switching phenomenon can be understood by analyzing the change of the spin susceptibility tensor, Eq. \eqref{spinsusceptibilty}, and the internal magnetic field, Eq. \eqref{currentMF}, due to the voltage pulse.
The periodicity shown in Fig. \ref{ripples} originates from the self-energy, Eq. \eqref{selfenergy}, where an applied pulse generates the phase factor $\exp\{i\varphi\}$ after the pulse is turned off. 
In Fig. \ref{interactions}(a) -- (d) we plot the integrated underlying fields for pulses of different temporal length, i.e., $\mathbf{j}(t)=\int \mathbf{j}(t,t')dt'$ and $\mathbb{J}(t) = \int \mathbb{J}(t,t')dt'$. It represents the first case of switching in Fig. \ref{ripples}(a) where the spin switches for $\varphi/2\pi=1.59$ and $3.34$.

The internal magnetic field in the z-direction, $j_{z}(t)$, is shown in Fig. \ref{interactions}(a). It illustrates rapid change immediately after the pulse is turned off and approaches a finite value in the long time limit.
The internal field gives mixed contributions depending on the voltage applied.
In the spin-flipping regime, $\varphi\in(2\pi, 4\pi)\mod{4\pi}$, exemplified by $\varphi/2\pi=1.59$ (blue)  and $3.18$ (black) in the figure, the field exhibits a drastic varying behavior and then reaches a constant value.
The drastic behavior occurs during the spin flip where the peak at $\varphi(t)/\varphi = 1.5$ is when $\dot{S_{z}}$ reaches its peak value.
In the spin-conserving regime, $\varphi\in(0,2\pi)\mod{4\pi}$, $\varphi/2\pi=2.39$ (red) in the figure, the changes in the field is less drastic and reaches about half the strength in the long time limit.
The significant change in the long time limit can be attributed to the change of the direction of the local spin moment as it is encoded in the GFs of the QD, cf., Eq. \ref{GF}. 

Considered as a self-interaction in the time-domain the Heisenberg interaction, $J_H$ is of anti-ferromagnetic character (positive) for all pulse lengths, see Fig. \ref{interactions}(b). Here, the change is not that significant for different pulse lengths although the time-evolution and the terminal value is clearly different in the two regimes.
The DM interaction $D_z$ changes sign in the spin-flipping regime, whereas it is strictly positive in the spin-conserving, see Fig. \ref{interactions}(c).
The Ising interaction includes both the dynamic contribution $\mathbb{I}_{zz}$ and the intrinsic uniaxial anisotropy D. The dynamic contribution is small but finite and it can easily be seen that the intrinsic contribution is dominating, see Fig. \ref{interactions}(d).
We also observe that the characteristics for $\varphi/2\pi=2.39$ is smaller by amplitude in comparison to the other pulses.
All fields depend strongly on the pulse length, bias voltage, temperature, magnetic field, exchange coupling and tunneling coupling.

A conclusion that can be drawn from the plots in Fig. \ref{interactions} is that within the spin-flipping regime, the induced interactions have a tendency to grow larger with increasing pulse length. The analogous behavior cannot, however, be observed by increasing the voltage bias and simultaneously decreasing the pulse length while preserving the phase $\varphi$. Although the non-linearity of the dynamical spin equation prevents us from determine the exact origin of this property, we conjecture that the different conditions leading to either conservation or flipping of the localized spin are not governed solely by the phase. It is rather a combination of the appropriate phase and that the time-evolution of the surrounding electronic structure accumulates density differently in the two cases.

Although the dominant fields in the transient dynamics are the Heisenberg interaction and the internal field, the anisotropic fields are crucial for the switching to occur. Due to the isotropic nature of the Heisenberg interaction, its corresponding potential landscape supports a degenerate set of stationary solutions for the spin, see left panel in Fig. \ref{potentials}(a). Hence, the stationary solution is always governed by the external field. While the degeneracy of the potential landscape is not broken by the Ising interaction and the intrinsic uniaxial anisotropy, it creates an energy barrier between the degenerate solutions, see right panel of Fig. \ref{potentials}(a). The height of this barrier effectively determines an upper boundary for the temperature in order to prevent thermal random drift between the two solutions. The DM interaction generates a spin transfer torque which, when sufficiently strong, can push the spin over the energy barrier, see Fig. \ref{potentials}(b). As retardation is inherent in the generalized SEOM by construction, both spin orientations, parallel and anti-parallel to the external field, constitute stable fixed points in the phase space of the dynamical system. Hence, the torque generated by the DM interaction merely has to be sufficiently large to push the system into the realms of the opposite solution for the switching to occur.  This is similar to the case where anisotropy is introduced in the system by magnetic leads of different polarization \cite{Hammar2016}. 

\begin{figure*}
	\includegraphics[width=\textwidth]{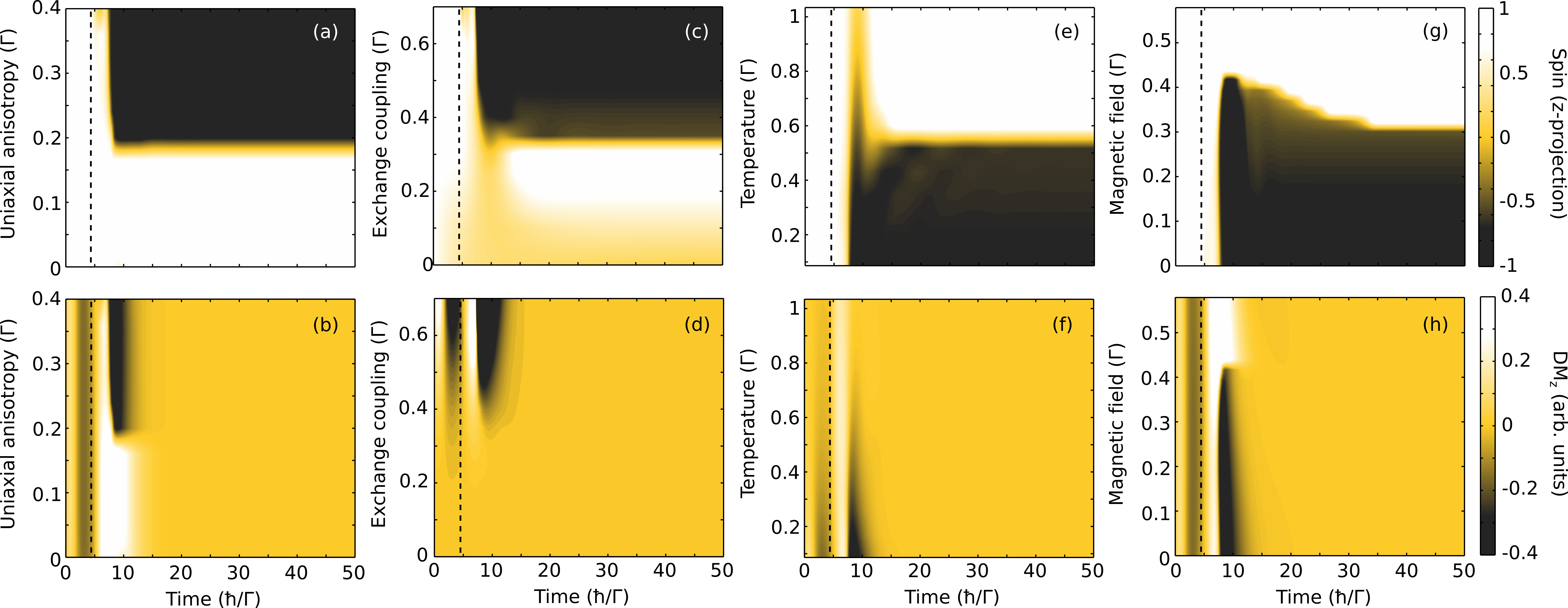}
	\caption{Resulting evolution of $S_{z}$ and corresponding change in the $\text{DM}_{z}$ field for varying (a, b) uniaxial anisotropy D, (c, d) different exchange coupling strength $v$, (e, f) temperature T and external magnetic field $B^{ext}$. Here a pulse $t_1-t_0 = 4.5 \hbar/\Gamma$ is applied and other parameters are hold constant with the same values as in Fig. \ref{ripples}. The vertical dotted line indicates when the pulse ends.}
	\label{diffparameters}
\end{figure*}

Tuning the DM interaction and the resulting spin transfer torque is of fundamental importance in order for the switching to occur. 
It is tuned by several competing parameters, e.g., intrinsic uniaxial anisotropy, local exchange, temperature, external magnetic field, and tunneling coupling to the leads. 
The intrinsic uniaxial anisotropy $D$ of the localized spin is required in order to create two separate ground states in the long time limit after the dynamic fields are switched off, cf., Fig. \ref{potentials}. This can be seen in Fig. \ref{diffparameters}(a), which shows the time evolution of the spin orientations for increasing anisotropy $D$ after a given pulse. The required anisotropy field needs to satisfy $D\gtrsim\Gamma/5$ in order to give a large enough barrier to overcome the thermal fluctuations. Fig. \ref{diffparameters}(b) shows the corresponding DM field in the z-direction for different uniaxial anisotropy and it can readily be shown that at $D\approx\Gamma/5$ the interaction changes sign, thus causing a switching by spin transfer torque.
Variations between the two stationary spin orientations are governed by the local exchange coupling $v$ between the spin and the electrons in the QD level. A local exchange integral satisfying $v\lesssim\Gamma/3$, does not sustain sufficiently strong transient internal fields to enable the switching. This can be seen in Fig. \ref{diffparameters}(c), which shows the time evolution of the spin orientations for increasing coupling $v$ after a given pulse. As the exchange integral satisfies $v\gtrsim\Gamma/3$, the spin undergoes a reorientation. This is also clearly illustrated by the DM field in Fig. \ref{diffparameters}(d) where there is first significant contributions above $v\gtrsim\Gamma/3$.

The switching is limited by the temperature and external magnetic field. From our simulations we can see that the limit on temperature $T$ and an effective spin switching requires that $Tk_B \lesssim\Gamma/2$, where $k_B$ is the Boltzmann constant, see Fig. \ref{diffparameters}(e). This happens as the temperature introduces thermal fluctuations to counteract the barrier between the two stable solutions, cf., Fig \ref{potentials},  and erases the dynamic features of the fields. It can be illustrated by the DM field in the z-direction for different temperatures where the negative features vanish, see Fig. \ref{diffparameters}(f).
Moreover, magnetic field strengths $g\mu_B B^{ext}\lesssim\Gamma/3$ is necessary for the spin switching since the induced fields  cannot overcome too strong external magnetic fields, see Fig. \ref{diffparameters}(g). It is clearly shown in the DM field that it changes sign when the spin no longer switches, see Fig. \ref{diffparameters}(h).

Regarding limitations in our approach we have not considered quantum spins or strongly correlated spins.
However, our model is essentially applicable for strongly localized spins, pertinent to, e.g., atomic transition metal and rare earth elements in molecular compounds such as phthalocyanines and porphyrins \cite{Chen2008, Wende:2007aa, Coronado2004, Urdampilleta2011}. Therefore, our model is restricted to large spin moments, for which a classical description is viable, while quantum spins are beyond our approach. We, moreover, assume the QD level to be resonant with the equilibrium chemical potential, hence, avoiding possible Kondo effect that otherwise may occur. While neglecting the local Coulomb repulsion is a severe simplification of the QD description, it is justified since it is typically negligible for the $sp$-orbitals that constitute the conducting levels in the molecular ligands structure.

Furthermore, we have not considered the effect of a thermal and random noise in the generalized SEOM. As motivated in Ref. [62] this requires that the energies of the interactions in the problem considered are larger than the energies of these thermal noise fields. Including such effects would add to the limitation of temperature already stated in the results.

\section{Conclusion}
In conclusion, we have demonstrated that phase induced switching of a localized magnetic moment embedded in a tunnel junction can be obtained for short voltage pulses $\tau$, satisfying $\varphi\in(2\pi,4\pi)\mod{4\pi}$.
The underlying rapid dynamics of the nanosystem and effects of memory are included through our newly developed generalized spin equation of motion procedure. The feedback of the spin onto itself through the surrounding environment is of vital importance as it provides a mechanism for a dynamical indirect electronically mediated spin-spin interaction. The switching phenomenon is also dependent on highly anisotropic transient fields, creating a pulse-dependent torque on the local spin.

\section{Acknowledgements}
This work is supported by Vetenskapsr\aa det and SNIC 2018/8-29.

\end{document}